*Title:*

# Practical considerations for crystallographic and microstructure mapping with direct electron detector-based electron backscatter diffraction

*Running Head:*

Practical considerations for DED-based EBSD

*Authors:*

Tianbi Zhang[1] (0000-0002-0035-9289), Ruth Birch[1] (0000-0003-1718-1568), Graeme Francolini[1] (0009-0007-2061-7744), Ebru Karakurt Uluscu[1] (0009-0002-2802-2677), Ben Britton[1*] (0000-0001-5343-9365)

1. Department of Materials Engineering, University of British Columbia

*Corresponding author: ben.britton@ubc.ca

*Conflict of Interest Statement:* Each of the authors declare no conflicts of interest.

*Graphical Abstract:*

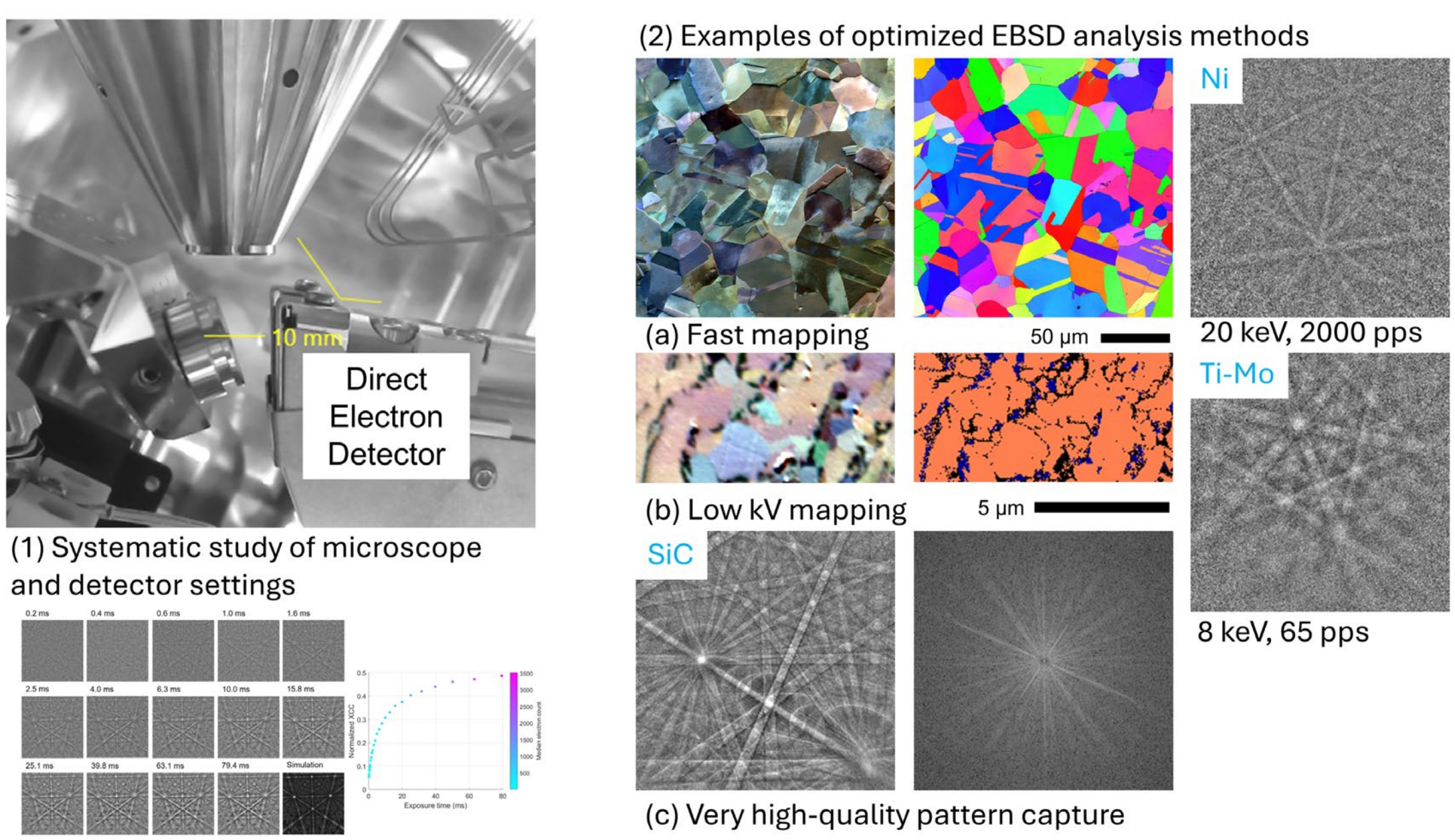

# Practical considerations for crystallographic and microstructure mapping with direct electron detector-based electron backscatter diffraction

Tianbi Zhang[1] (0000-0002-0035-9289), Ruth Birch[1] (0000-0003-1718-1568), Graeme Francolini[1] (0009-0007-2061-7744), Ebru Karakurt Uluscu[1] (0009-0002-2802-2677), Ben Britton[1*] (0000-0001-5343-9365)

1. Department of Materials Engineering, University of British Columbia

*Corresponding Author: ben.britton@ubc.ca

## Abstract

Compact direct electron detectors are becoming increasingly popular in electron microscopy applications including electron backscatter diffraction, as they offer an opportunity for low cost and accessible microstructural analysis. In this work, we explore how one of these commercial devices based on the Timepix chip can be optimized to obtain high quality data quickly and easily, through careful systematic analysis of a variety of samples, including: semiconductor silicon, commercially pure nickel, a dual phase titanium-molybdenum alloy, and a silicon carbide ceramic matrix composite. Our findings provide strategies for very fast collection of orientation maps, including at low voltage (5-10 keV) and low beam current conditions. Additionally, strategies for collection of very high quality EBSD patterns are demonstrated that have significant potential for advanced EBSD applications (e.g. elastic strain mapping).

## Introduction

Electron backscatter diffraction (EBSD) is a popular orientation microscopy technique in the scanning electron microscope (SEM) (Schwartz et al., 2009) and is most commonly used to resolve features in micro- and nanometer scale in crystalline materials over a large area (a few micrometers to millimeters) through capturing and analyzing backscattered Kikuchi patterns over a grid of positions (Wilkinson & Britton, 2012). This technique has seen considerable development since it was first demonstrated in an SEM by Venables et al. in the 1970s (Venables & Bin-jaya, 1977; Venables & Harland, 1972), in terms of pattern processing and analysis routines for automation and on-line analysis (Dingley et al., 1987; Krieger Lassen et al., 1992; Chen et al., 2015), diffraction physics studies (Deal et al., 2008; Winkelmann et al., 2007, 2010: 201), experimental geometries (Thaveeprungsriporn & Thong-Aram, 2000; Mingard et al., 2018; Zhang & Britton, 2024) and more recently the use of digital direct electron detectors (DED) (Wilkinson et al., 2013; Vespucci et al., 2015; Mingard et al., 2018; Alex Foden et al., 2019; Wang et al., 2021; DeRonja et al., 2024) to routinely capture higher quality patterns.

The main advantages of using DED for EBSD experiments, as compared to conventional indirect scintillator-based CCD or CMOS detectors include:

(1) Improved dose efficiency and signal-noise ratio (SNR), characterized by high detector quantum efficiency (DQE). This is a result of the higher efficiency of direct electron detection, and much lower noise from event (electron) counting.
(2) Optical distortion free diffraction patterns as no scintillator or optical coupling is involved (c.f. the distortion effects shown in (Dingley, 2004)).
(3) Potential energy filtering of diffraction patterns to further improve pattern quality.

These advantages have been clearly established in the founding works of this area, including work by Wilkinson et al. (2013) and Vespucci et al. (2015) that focused on acquisition of single patterns rather than scanning. The high-quality patterns shown in these works were obtained at exposure times of 10 s, 50 s and 1482 s (24.7 min), which are not practical for efficient microstructure mapping. A more practical exposure time (0.095 s) was used by DeRonja et al. (2024) for comparing five commercial EBSD detectors including a DED to evaluate patterns for cross-correlation based analysis such as HR-EBSD. Even shorter exposure times can be used for band detection-based pattern indexing (e.g. Hough or Radon transform-based methods) as the band detection methods have been developed for lower signal to noise patterns and still enable reliable online pattern indexing and crystal phase/orientation determination. For example, Zhang & Britton (2024) reported indexable patterns from Si (001) crystal with an exposure time of 0.02 s. Wang et al. (2021) demonstrated microstructure mapping of a Ni alloy at 281 patterns per second (pps) (i.e., ~0.0036 s/pattern) using a monolithic active pixel sensor (MAPS), and they obtained 5988 pps mapping using sparse sampling of pixels together with inpainting methods. In practice, the rate of collection of indexable patterns that is achieved is a combination of the electron counter saturation level, exposure time, detector readout rate and the number of electrons that form the diffraction pattern, and for instance Wang et al. used ~20x the beam current of Zhang and Britton.

Another application of DEDs for EBSD is for low primary beam energy (5-10 keV) and/or low beam current (<1 nA) measurements on beam sensitive material, where indirect detectors struggle to capture high quality patterns due to high noise and low scintillator efficiency. This was first demonstrated by Adhyaksa et al. (2018) using a Medipix chip-based DED and an exposure time of 0.1 s on halide perovskite specimens at 30 keV primary beam voltage and 100 pA beam current, and Muscarella et al. (2019) studied perovskite thin films at 15 keV, 100 pA, 0.1 s exposure, Wang et al. (2021) studied Si at 4-10 keV and Ni at 10 keV, using a 32 nA probe current and (Della Ventura et al., 2025) studied $HfO_2$ and $HfSiO_4$ at 5-10 keV, 1.6-3.2 nA and exposure times of 20 ms (10 keV mapping), 67 ms (5 keV mapping) and 4 s (individual pattern).

This literature review indicates that the advent of wider access to these technologies motivates a study to explore the optimum conditions that can be used to obtain high quality EBSD data, as well as strategies to achieve these. This present study explores the interplay between beam current, detector energy threshold and exposure time and their effects on pattern quality, to collect data efficiently for different types of EBSD experiments using DEDs. A major outcome of our work highlights that pattern quality is ultimately determined by the number of “useful” electrons (i.e. electrons that contribute to the Kikuchi band features which are used for indexing and microstructural analysis) captured by the detector. This means that there are strategies that can be applied to optimize an experiment based upon the microscope set up and detector collection parameters to help optimize pattern quality and/or acquisition speed. Based on the parametric study using individual Si patterns, case studies of microstructure mapping of different polycrystalline samples: commercially pure Ni, Ti-Mo alloy (with an α-Ti HCP matrix and a Mo-rich BCC β-ligament interface phase), SiC ceramic matrix composite (with the 4H and 6H polymorphs (Bind, 1978)) are also presented to explore multi-phase and lower symmetry materials which may create indexing challenges.

## Methods

EBSD experiments were performed on a Thermo Fisher Scientific Apreo S Chemi SEM with a TruePix EBSD detector. A schematic of this hybrid pixel detector and a chamberscope image of the EBSD set up is presented in Figure 1. The sample is tilted 70° towards the detector, whose sensor is parallel to the primary electron beam. A 110 V bias was applied to the SEM sample stage.

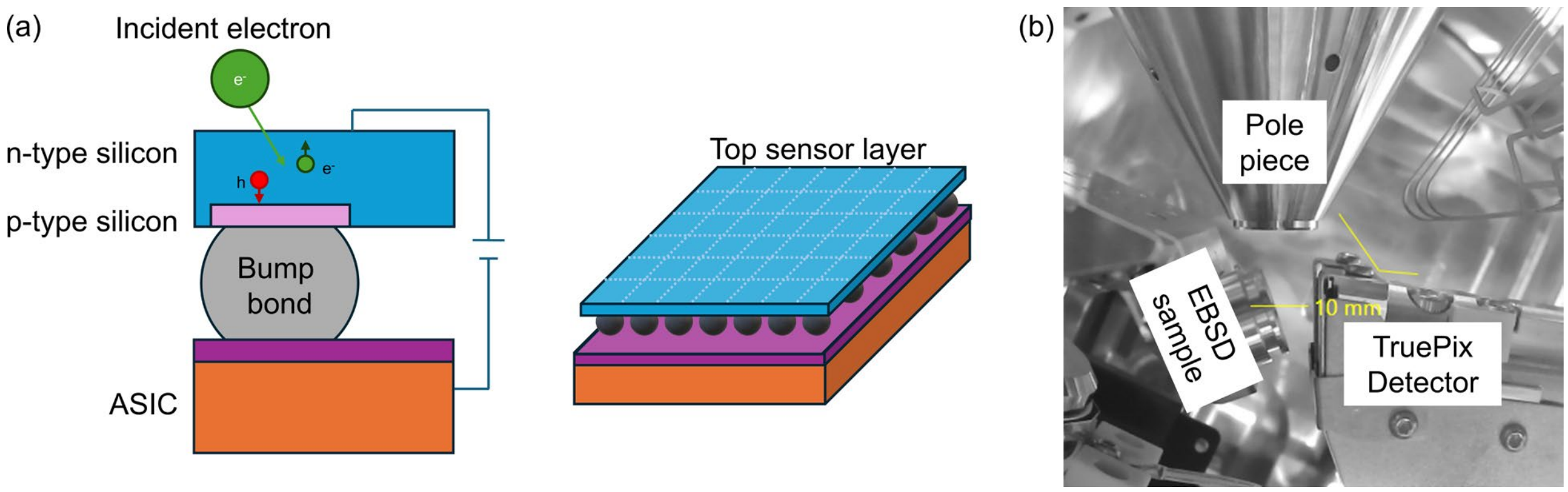

Figure 1: (a) schematic of a hybrid pixel detector (HPD), which consists of a sensor layer of single crystal silicon that can detect and count incident electrons, via the generation of electron-hole pairs within a semiconductor structure. Pixel resolution in the chip is achieved by the bump bonding of electrical read out contacts onto an application-specific integrated circuit (ASIC); (b) Chamber scope image of the EBSD setup, including the TruePix (HPD).

The TruePix detector is a HPD DED based on the Timepix chip (Llopart et al., 2007) and its control is integrated with microscope workflow. The detector uses a single crystal Si-sensor layer which is bump-bonded onto the readout chip with 256 x 256 pixels (55 µm pitch), for a total sensor size of ~14 x 14 mm. The detector counts the number of electrons exceed the energy threshold setting of the detector (which can be set between the software minimum of 3.5 keV and the energy of the primary electron beam). The counter is 14-bit (0-11810 counts), and electron counts are used to reconstruct the raw diffraction pattern. The median of the raw diffraction pattern, “median electron count” (MEC), is used as a metric of electron dose received by the detector. A reverse bias of 50 V is applied to the sensor of the detector to achieve full depletion. To collect diffraction patterns, the detector operates in frame readout, which allows a minimum exposure time of 0.1 ms. There is an additional ~300 µs per pattern, and so in the absence of very fast beam blanking the total dwell time is the exposure time plus 300 µs.

Initial calibration of the detector enables the use of a software-based routine to determine the maximum allowed exposure time for the current field of view (FoV) of the microscope on the sample, as well as microscope and detector parameters (e.g. primary beam voltage, beam current, energy threshold) and this minimizes saturation of the pixels. For this work, exposure times do not exceed this maximum and every pattern is a single frame captured each time. Within the software, patterns can be flat fielded against an average "pattern" collected from a large area and/or many different grain orientations as an approximation of the diffuse background to improve pattern contrast for indexing. Raw patterns are stored as 16-bit integers and corrected patterns as double precision floating-point numbers. In addition, microstructural imaging is realized using virtual detectors (subregions on the detector) from the top (black-white contrast) and bottom (red-green-blue contrast using three subregions) during online pattern processing and indexing.

Collection and online processing of EBSD data were performed in the Thermo Fisher xTalView software (v 1.0.0.1139). This involves the classic Radon transform, band localization, band indexing, crystal orientation determination, and phase identification, which is performed concurrently with pattern acquisition. Dynamical EBSD patterns of Si were simulated using the Bloch wave approach by Winkelmann et al. (Winkelmann et al., 2021) to facilitate direct pattern comparison of Si patterns as a measure of quality of experimental patterns using normalized cross-correlation within MATLAB, with pattern matching and reprojection performed using functions within AstroEBSD (Britton et al., 2018).

## Results and Discussions

The structure of this section is as follows: (a) systematic studies of the effects of exposure time, energy threshold, beam current, and median electron counts on pattern quality using the Si (001) crystal sample; (b) case studies on microstructural mapping of three samples (Ni, Ti-Mo alloy, SiC); (c) examples of very high quality patterns from Si and SiC to highlight the potential of this detector with regards to advanced analysis methods.

**Effect of exposure time.** Figure 2 shows EBSD patterns of Si (001) captured at 20 keV, 3.2 nA beam current, 11 keV energy threshold, and exposure times from 0.2 ms to 80 ms. A visually discernable pattern can already be observed at 0.4 ms exposure time, and pattern sharpness gradually increased with exposure time. This is also confirmed with a measure of pattern quality by comparing the experimental patterns with a dynamically simulated pattern matched with the longest-exposed pattern. The shortest-exposed pattern gives a low normalized cross-correlation coefficient (XCC) of only 0.05, and XCC rapidly increases with exposure time. This supports the general idea that more electrons captured results in higher pattern quality. The maximum achievable XCC starts to show at even longer exposure times, where the signal to noise seems to saturate. Note that an XCC of 1 is unlikely due to simplifications used to generate the dynamical pattern simulations.

For design of experiments, exposure time can be selected by balancing the requirements of pattern quality (e.g. to achieve 80% of the maximum XCC) and acquisition speed. In addition, pattern quality may be sufficient at even shorter exposure times for line detection-based indexing. For example, the 1 ms pattern was also correctly indexed by Thermo Fisher xTalView.

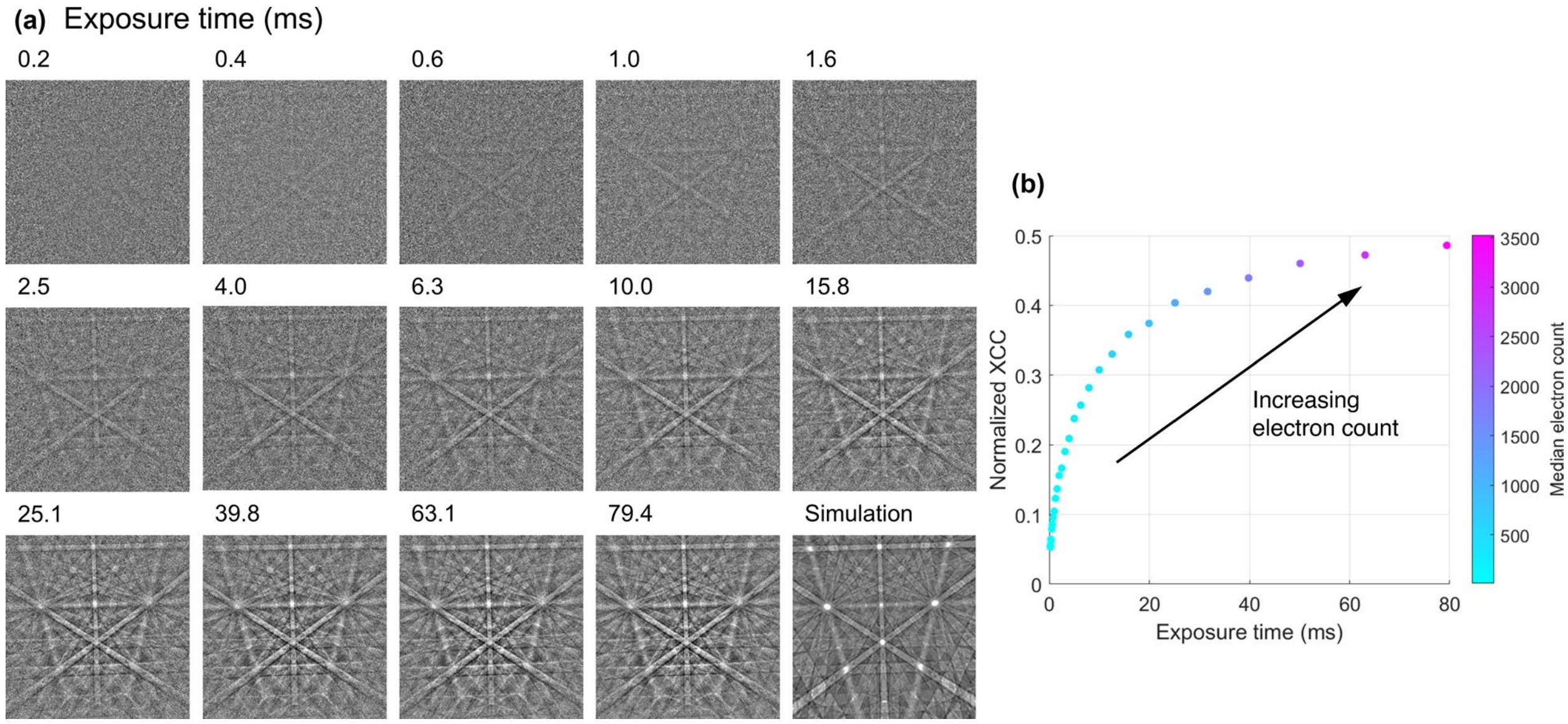

Figure 2. (a) Si (001) EBSD patterns (background corrected) captured at 20 keV primary beam energy, 3.2 nA beam current, 11 keV detector energy threshold and different exposure times. Matched dynamically simulated pattern is included for reference. (b) normalized cross-correlation coefficients between experimental and matched, dynamically simulated pattern as a function of exposure time and median electron count. (Color key represents median electron count).

**Effect of energy threshold.** Figure 3 shows EBSD patterns of Si (001) captured at 20 keV, four levels of beam currents, nine levels of energy thresholds, and 10 ms exposure time. The fraction of electrons of the primary electron beam captured by the EBSD detector as a function of energy threshold follows the same relationship at different beam currents except for the lowest energy threshold, suggesting that the counting characteristics of the detector should be consistent at these different incident electron dose rates. The behavior at the lowest energy threshold may be related to stray electrons and higher extent of charge sharing due to an incident electron being detected by multiple pixels. This can happen as the incident electron can scatter within the detector and make electron hole pairs that are collected by neighboring pixels.

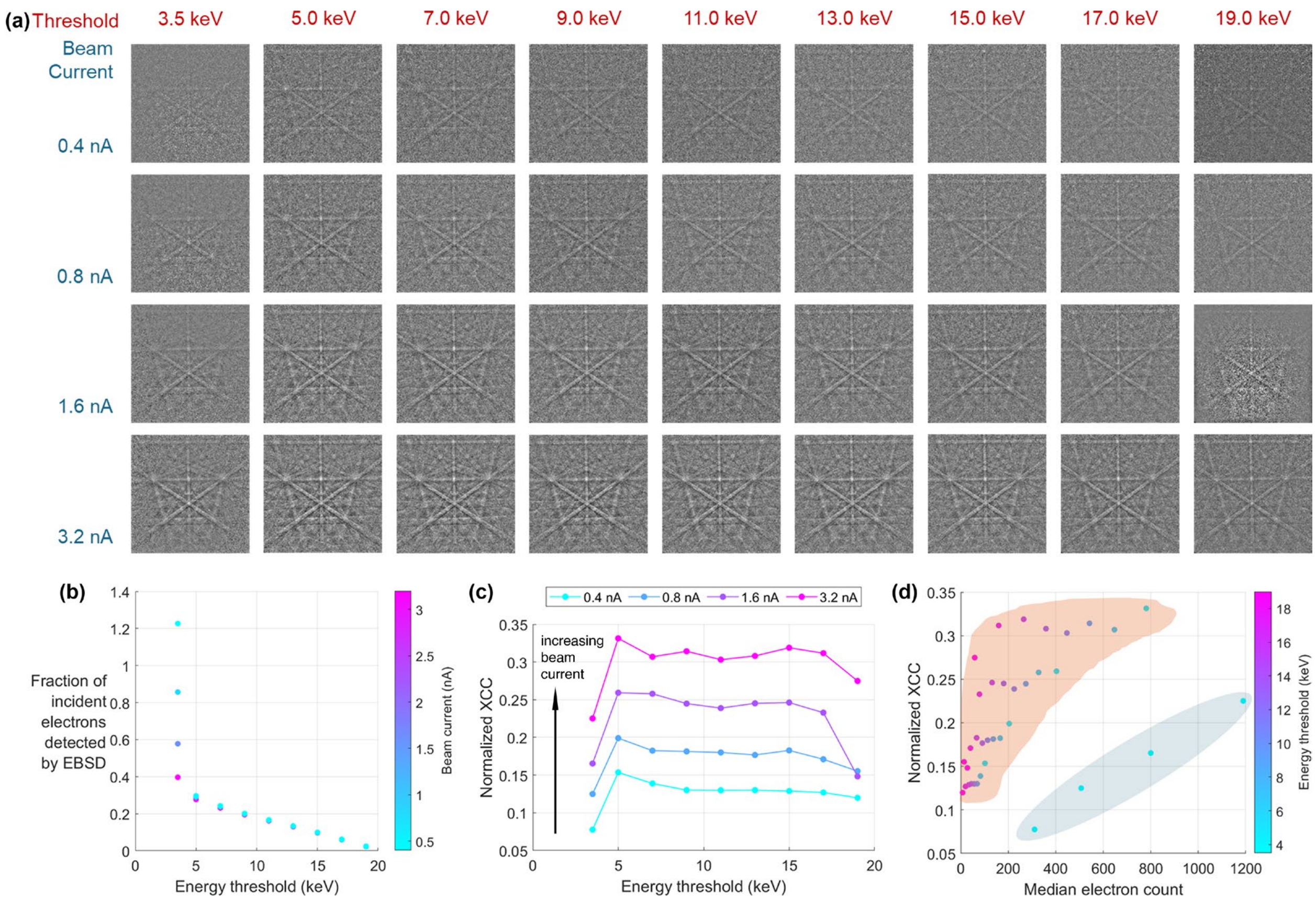

Figure 3. (a) Raw Si (001) EBSD patterns captured at 20 keV primary beam energy, three different beam currents, nine detector energy threshold and 10 ms exposure time. (b-d) Normalized cross-correlation coefficients between experimental and the matched, dynamically simulated pattern, as well as median electron count (MEC), as a function of energy threshold. The shaded orange region on (d) indicates conditions where the energy filter improves the pattern quality significantly. (Color key represents beam current in (b) and energy threshold in (d)).

For a primary beam energy of 20 keV and at the same beam current, XCC plateaus in the 5-17 keV threshold range, and reduction of the number of electrons captured by energy filtering only has a minor effect on pattern quality until 19.0 keV (95% of the primary beam energy), which results in a reduction of pattern quality as the total number of electrons counted is very low. This result also aligns with the general observation that the DQE decreases at higher energy thresholds (e.g. (McMullan et al., 2007; Moldovan et al., 2012; Matheson et al., 2017)), and the effect of energy filtering on improving the modulation transfer function (MTF, related to pattern sharpness) is seen more for longer exposure times (e.g. Figure 2).

When a high threshold is used, the beam current and/or exposure time can be adjusted to compensate for the reduced electrons detected. The HDP can directly count the total number of electrons per pixel, reported here as the median electron count (MEC). This metric can be used to meaningfully adjust the microscope and director settings and improve pattern quality. An example of this is shown in Figure 4 where patterns of similar MEC values (240-260) were captured from Si (001) at 20 keV and different combinations of beam current, energy threshold and exposure time.

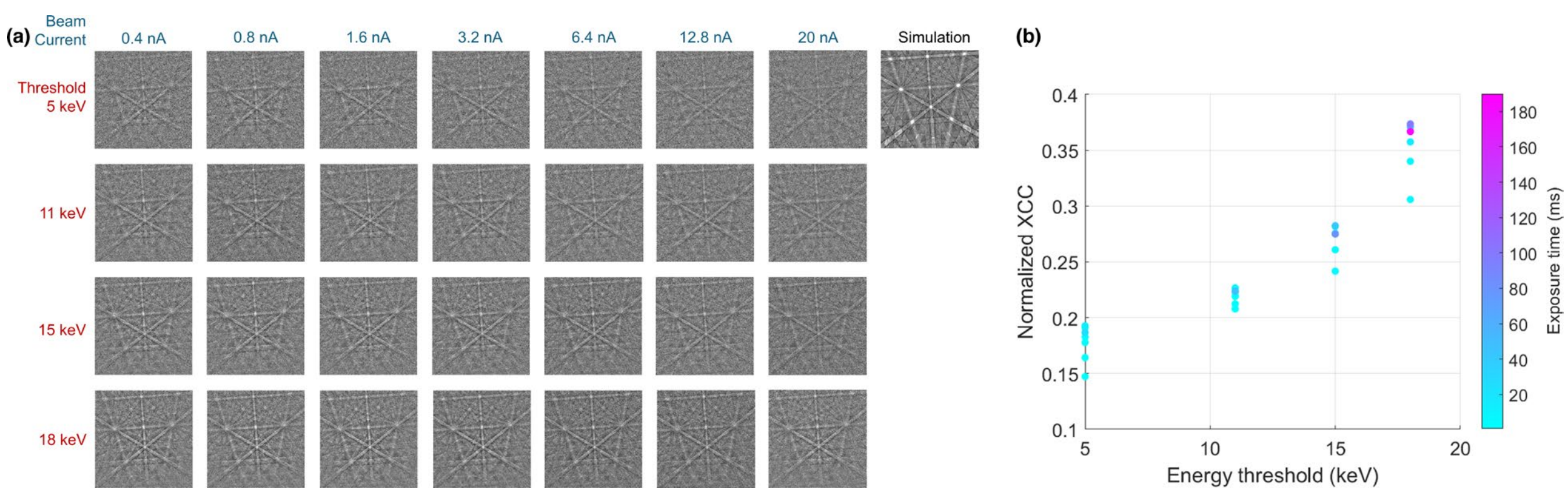

Figure 4. (a) Si (001) EBSD patterns captured at 20 keV primary beam energy and different combinations of beam current, energy threshold and exposure time, so that all patterns have a median electron count within 240-260. (b) Normalized cross-correlation coefficients between experimental and matched, dynamically simulated pattern, as a function of energy threshold (color scale represents exposure time).

Figure 4 shows that with equivalent number of electrons counted, there is an improvement of pattern quality when the energy threshold is increased, and this emphasizes that the “useful dose” (low energy loss electrons captured that contribute to the Kikuchi bands) captured is more important than the total dose captured. This result strongly agrees with prior energy filtering studies (Deal et al., 2008; Vespucci et al., 2015). In practical terms, this means fast mapping can be performed with an increase in probe current to offset the reduction in total electron dose detected due to the high energy filter.

**Low-keV performance.** Similar parametric studies at lower primary electron beam energies (5 and 10 keV), 10 ms exposure time and different energy thresholds and beam currents are presented in Figure 5 and 6.

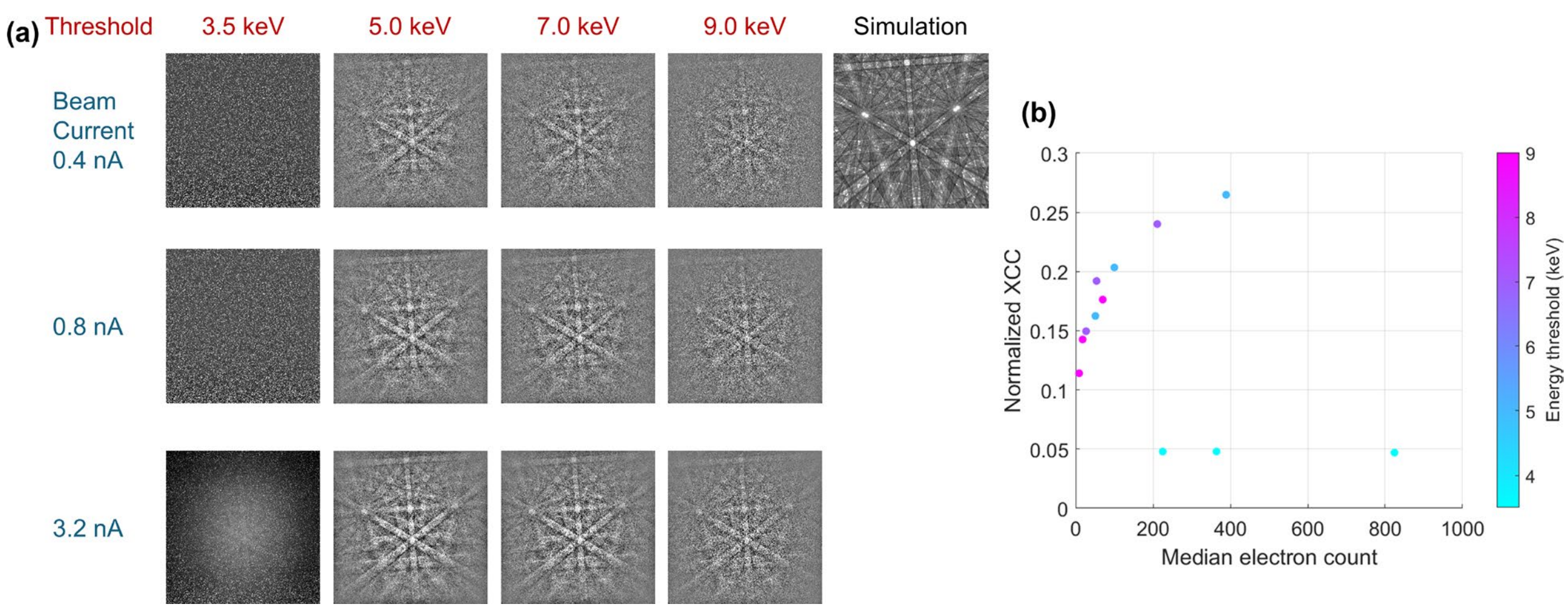

Figure 5. (Left) Si (001) EBSD patterns captured at 10 keV primary beam energy, three different beam currents, four detector energy threshold and 10 ms exposure time. (Right) Normalized cross-correlation coefficients between experimental and matched, dynamically simulated pattern, as a function of MEC and energy threshold. (Color key represents energy threshold).

At 10 keV, the highest XCC is obtained at 3.2 nA and 7.0 keV threshold. While a similar MEC can be obtained for 0.8 nA and 3.5 keV threshold, the captured pattern does not contain as strong Kikuchi bands and has a poor XCC value. This again aligns with the “useful dose” observation. Furthermore, patterns captured at 9 keV threshold are again affected by the reduced MEC. The DED also operates reasonably for a 5 keV primary beam energy. Thus, similar to the 20 keV study, pattern quality can be increased significantly through compensation of the reduced MEC by increasing the exposure time and/or beam current. Here we note that the primary beam energy is close to the minimum energy threshold value (and noise floor) of the detector, and this can cause reduced pattern quality when the energy threshold is low.

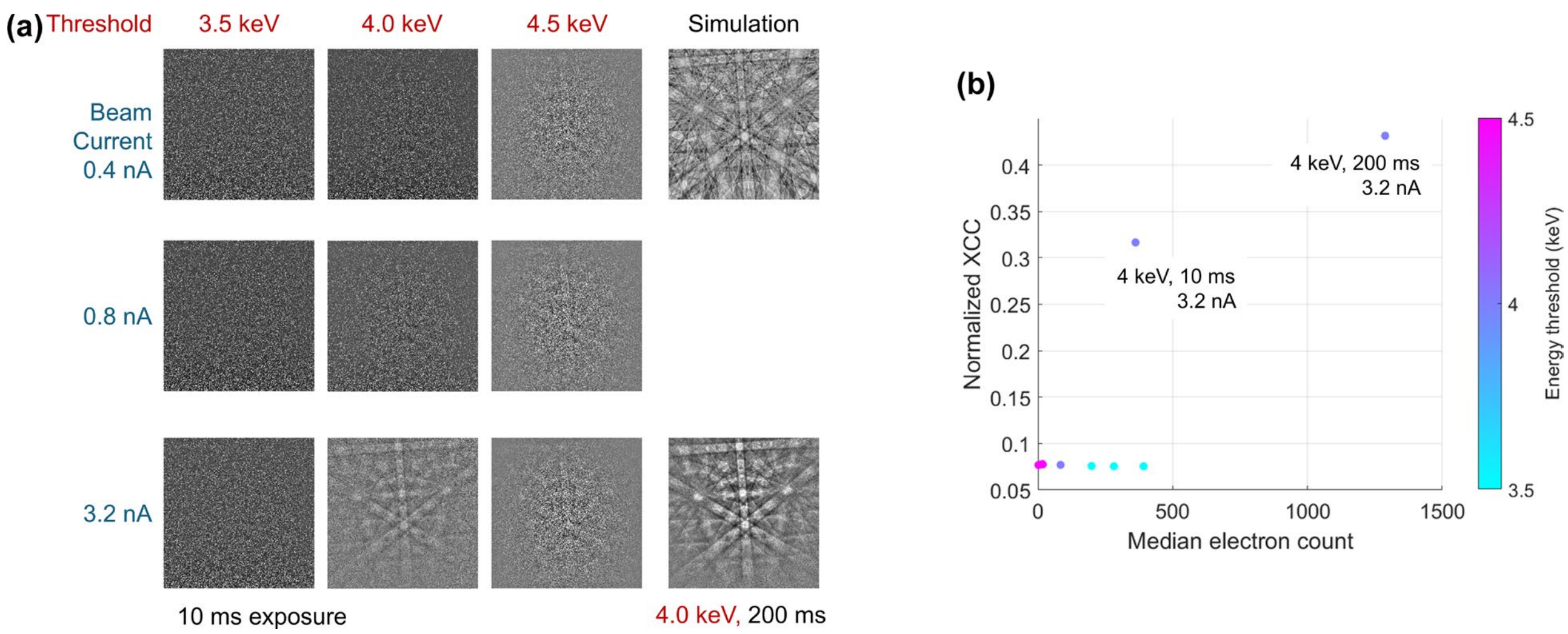

Figure 6. Raw Si (001) EBSD patterns captured at 5 keV primary beam energy, three different beam currents, four detector energy threshold and 10 ms exposure time. For reference, a higher-quality pattern captured at 200 ms exposure time is included. (Color key represents energy threshold).

### Case studies on microstructure mapping

The systematic study of single patterns of Si provides some guidance on the optimal conditions for pattern collection and microstructural mapping. This motivates sharing of case studies of 'typical' materials and for 'typical' EBSD analysis of microstructure mapping.

**Case Study 1: Nickel**

Figure 7 shows mapping of the Ni alloy at three primary beam energies and different combinations of exposure time and energy threshold. This example shows that by adjusting the energy threshold and exposure time, comparable indexing quality can be achieved at both high (20 keV) and low (5-10 keV) primary electron beam energies. A slower scan can capture high quality patterns which would likely be more suitable for more advanced EBSD analysis (e.g. high angular analysis), and a much faster scan can be performed at reduced energy thresholds and 2-10 times shorter exposure time.

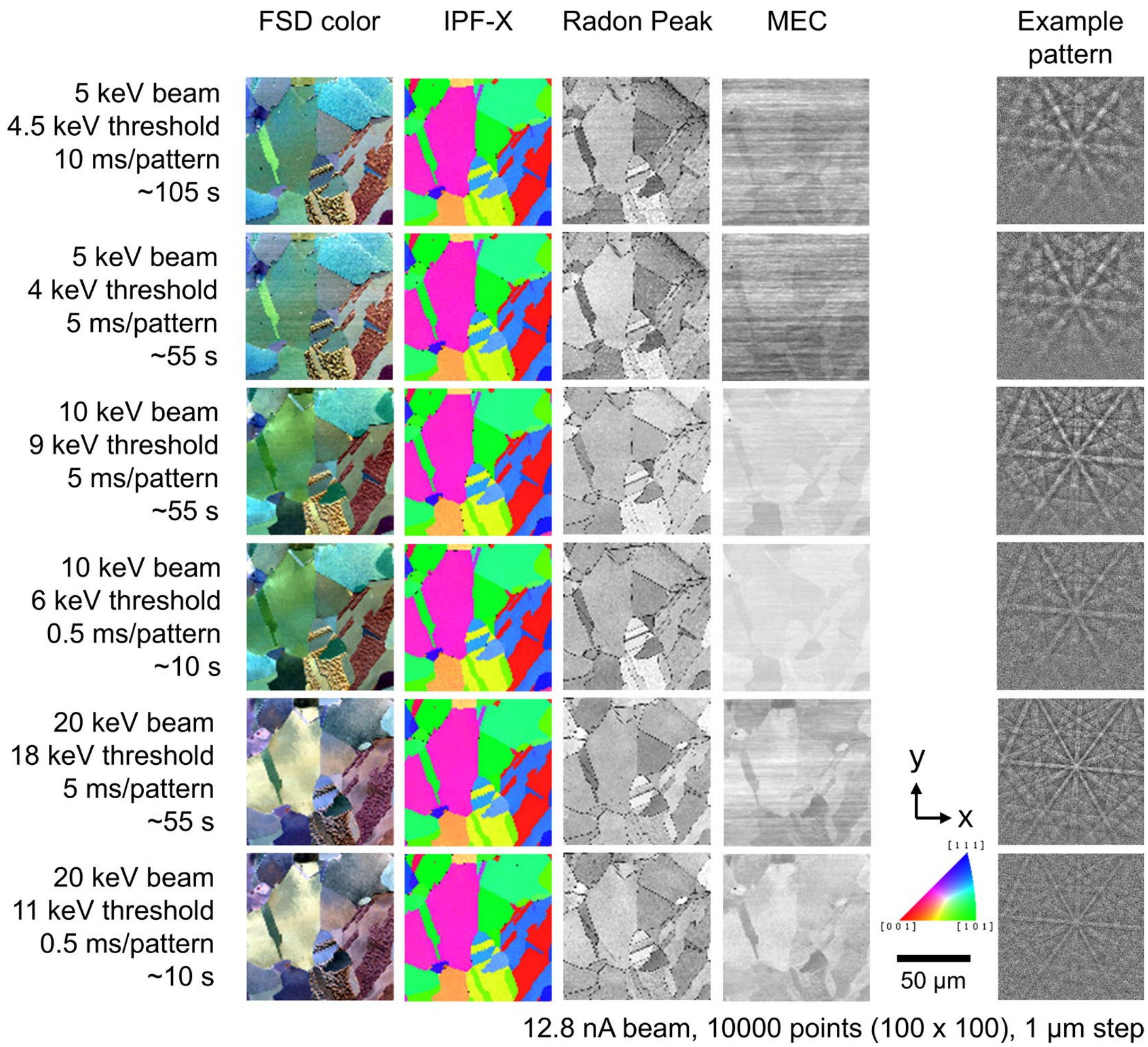

Figure 7. Microstructure map of Ni collected at 5, 10 and 20 keV primary beam energy, 12.8 nA beam current and different detector parameters to achieve faster acquisition time or higher pattern quality. Forescatter diode (FSD) color map, crystal orientation as represented with the IPF-X color key, example patterns, the pattern quality as measured via the relative intensity of the bands used to index the pattern from the Radon transform and the median electron count (MEC). Maps are 100x100 points with 1 μm step size. (color version online)

The faster pattern capture method can be exploited to collect maps that contain many more analysis points, and Figure 8 shows mapping of Ni alloy at the fastest speed possible for this current system, using a pattern exposure time of 0.2 ms and a 0.3 ms readout delay. This results in a total pattern collection time of 0.5 ms / pattern, i.e. collection at 2000 pps, which is comparable to many currently used indirect detectors (DeRonja et al., 2024). Horizontal strips can be seen on the MEC maps, which may be related to beam instability during the scan in terms of the probe current.

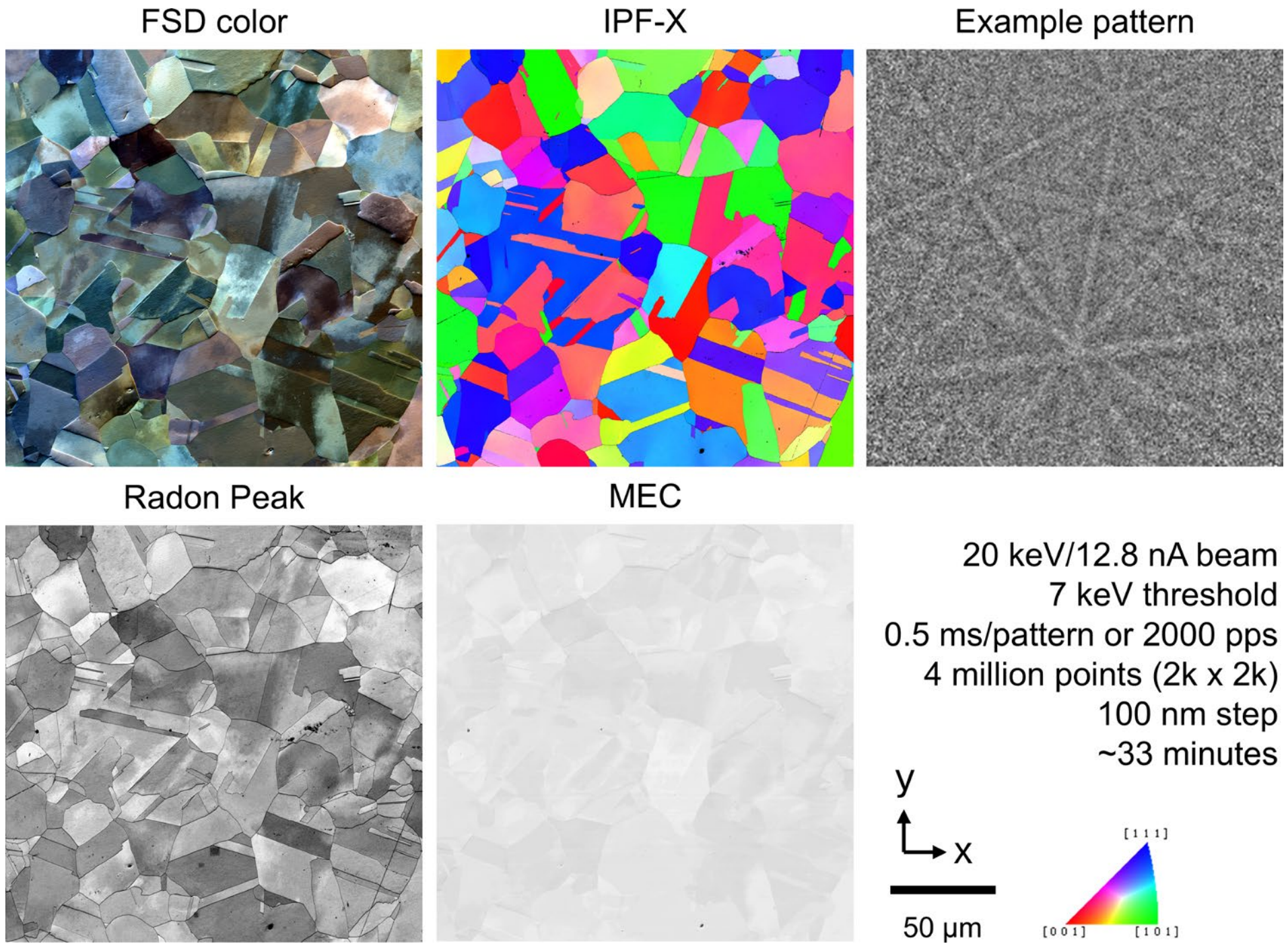

Figure 8. Microstructure map of Ni collected with a 2000 pps EBSD pattern collection rate. Forescatter diode (FSD) color map, crystal orientation as represented with the IPF-X color key, an example pattern, the pattern quality as measured via the relative intensity of the bands used to index the pattern from the Radon transform and the median electron count (MEC). This map is 2000 x 2000 points, with 100 nm step size. (color version online)

**Case Study 2: Ti-2wt%Mo**

Next, the more complicated Ti-2wt%Mo sample is explored at three conditions and are shown in Figure 9. The grain boundary regions, which contain very thin β-ligaments, are not easily indexed. These regions are however revealed within the FSD color micrographs and the MEC maps, and the MEC maps show that the Mo-rich β-ligaments increase the total electron dose that strikes the EBSD detector, akin to the Z-number contrast shown in conventional backscatter electron micrographs. Note that indexable patterns cannot be obtained at primary beam energy below 8 keV, and this is likely due to the presence of a thin surface oxide layer (2-20 nm) that always forms on titanium alloys when exposed to air or even a small partial pressure of oxygen. Difficulties in indexing the β-phase can be the result of topographical shadowing of the resulting EBSD pattern, which reduces pattern contrast and limits indexing success, as well as features that are too small for these beam conditions to generate a sufficiently indexable diffraction pattern that can be classified as one phase or another.

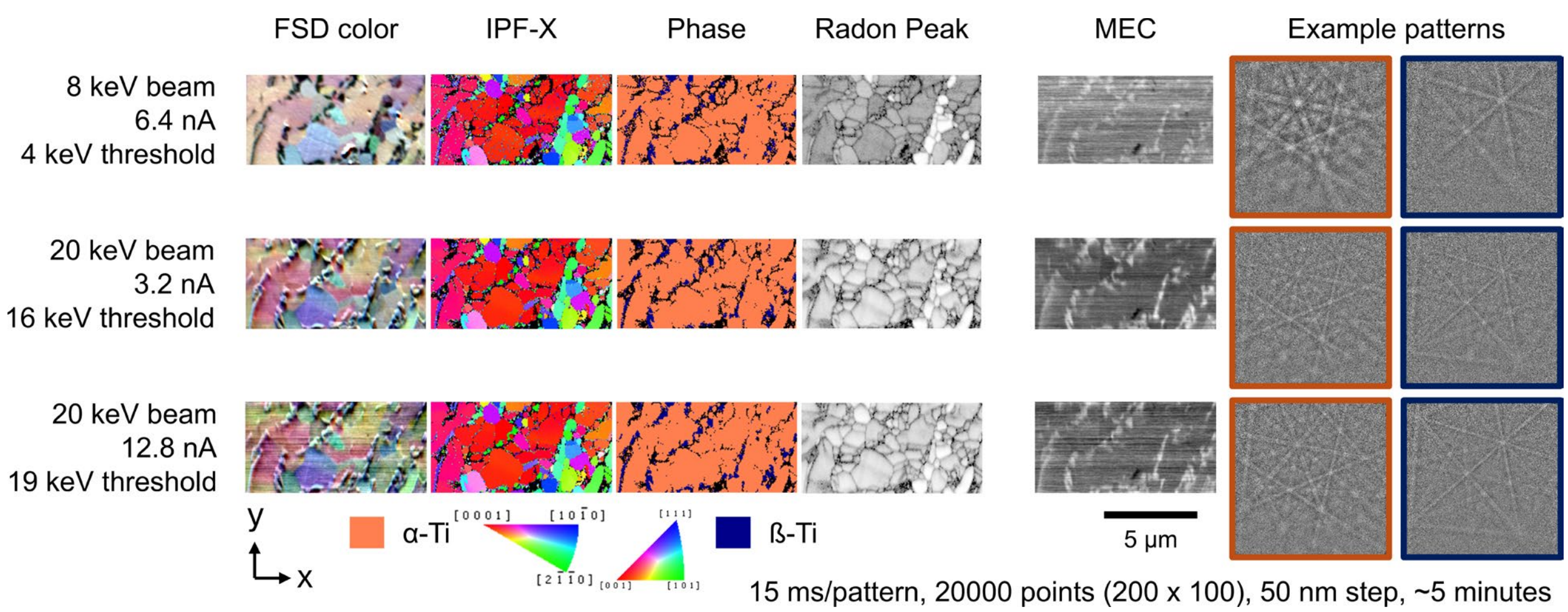

Figure 9. Microstructure maps of the Ti-Mo alloy collected at 8 and 20 keV primary beam energy, different beam currents and detector parameters. Forescatter diode (FSD) color map, crystal orientation as represented with the IPF-X color key, example patterns, the pattern quality as measured via the relative intensity of the bands used to index the pattern from the Radon transform and the median electron count (MEC). Maps are 200 x 100 points with 50 nm step size. (color version online)

**Case Study 3: SiC**

Next, the SiC sample is mapped at three primary beam voltages with optimized microscope and detector parameters and the results are shown in Figure 10. While the patterns for low keV are of high quality, reliable indexing is more difficult, and this is most likely due to the fact that the Radon peak detection algorithms have typically been tuned to analyze sharp and narrow bands more typical for higher primary beam energies.

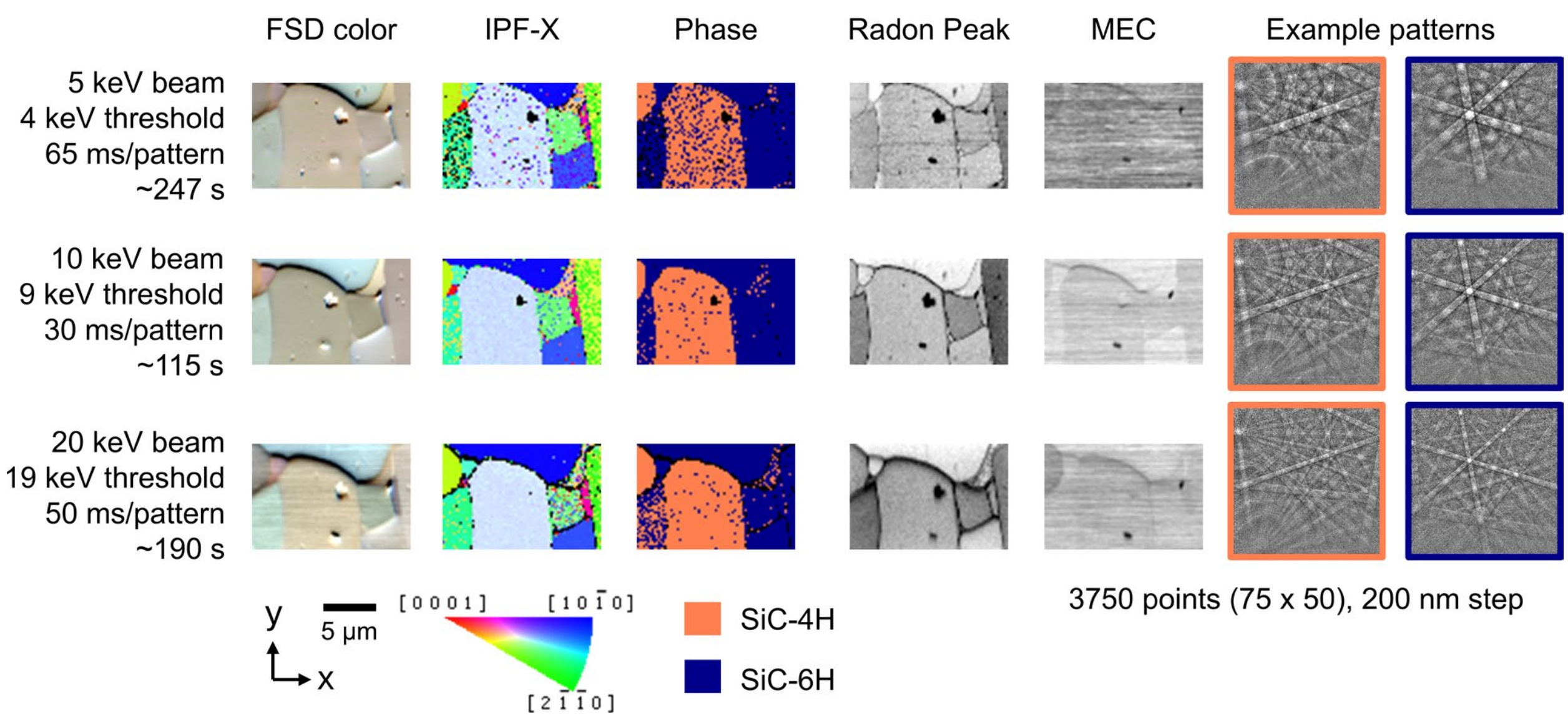

Figure 10. Microstructure maps of the SiC sample- collected at 5, 10 and 20 keV primary beam energy, 12.8 nA beam currents and different detector parameters. Forescatter diode (FSD) color map, crystal orientation as represented with the IPF-X color key, example patterns, the pattern quality as measured via the relative intensity of the bands used to index the pattern from the Radon transform and the median electron count (MEC). Maps are 75 x 50 points with 200 nm step size. (color version online)

**Capture of high-quality patterns.** Similar to the works by Wilkinson et al. and Vespucci et al. (Wilkinson et al., 2013; Vespucci et al., 2015), Figure 11 shows two examples of very high quality EBSD patterns from Si (001) and SiC (4H polymorph) using high energy filtering and long exposure times. These patterns include sharp higher order Laue zone (HOLZ) rings and higher order bands. The patterns are presented alongside dynamically simulated patterns of each crystal, and their log-power 2D fast Fourier transform spectra.

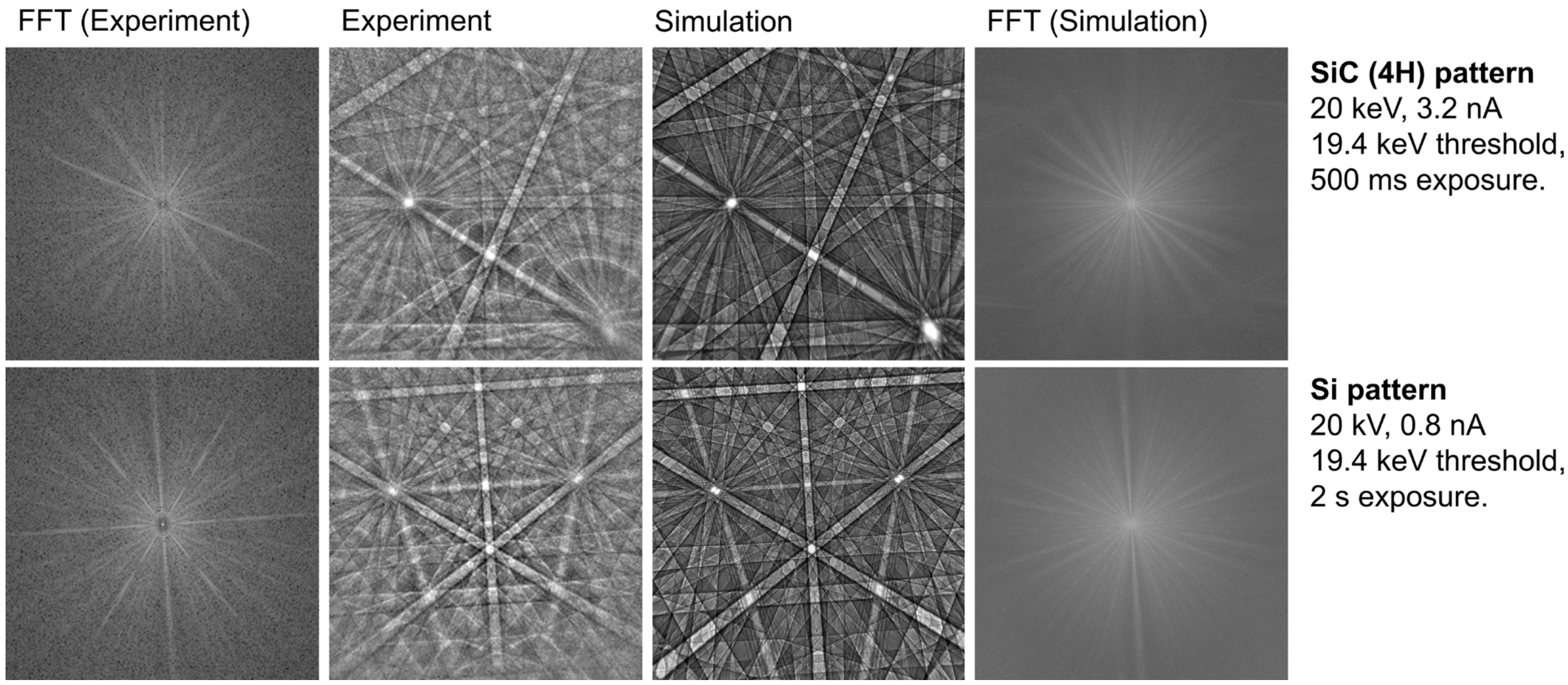

Figure 11. High quality EBSD patterns from Si (001) and SiC (4H polymorph), matched dynamically simulated patterns and power spectra.

## Discussion

This study highlights that a DED can be tuned to optimize pattern collection for an EBSD experiment. A few notable, basic elements of the experiment are considered as the following:

(1) The detector should be optimized to emphasize the ‘useful’ electrons captured for higher quality EBSD patterns. Optimization of the experiment can include tuning of (a) the primary beam energy, (b) the energy threshold of the detector; (c) the beam current; (d) the exposure time.

(2) The primary beam energy determines the width of the Kikuchi bands and the size of the electron probe in the sample, most notably the depth and scattering width of the interaction volume. A higher energy is often favorable for current indexing algorithms that use line-indexing based upon the Hough or Radon transforms, and additionally to penetrate through surface oxide layers or coatings. A lower voltage may be advantageous for beam sensitive samples and for improving spatial resolution (Steinmetz & Zaefferer, 2010). The primary beam energy must be above the minimum energy threshold of the detector.

(3) The energy threshold of the detector can be optimized to increase the “useful electron count” for patterns captured. A threshold above ½ of the primary electron beam can be used for fast pattern indexing, and higher thresholds can be used to increase the acuity and quality of the detected pattern at comparable numbers of electrons detected.

(4) The beam current can be increased to increase the total electrons that strike the detector per unit time. This is important when higher detector thresholds are used, as a significant fraction of incident electrons will be rejected by the detector. Note that typical electron microscopes have large spot sizes at higher beam currents, and sample degradation may occur at high doses and high dose rate (e.g. contamination, sample heating, or degradation of the crystal). Lower beam currents can be employed together with a lower threshold for more beam sensitive samples.

(5) The exposure time can be increased to collect more electrons, up to saturation of the electron counter, or reduced to favor fast experiments (e.g. *in situ* or 3D experiments) or mapping large areas to collect statistically useful data. Longer exposure times can be risky if there is a significant sample or microscope drift, and other strategies to drift correct may need to be employed (e.g. Tong & Ben

Britton, 2021). Note that the detector readout time must also be factored into the total dwell time per point in the EBSD scan.

In addition to the demonstrated success of online indexing approaches in this work, raw and processed patterns can be used for offline analysis, especially pattern matching methods, to improve indexing especially for low-symmetry structures. In line with Wilkinson et al. (2013) and Vespucci et al. (2015), very high-quality EBSD patterns can be obtained with the present set-up using high energy threshold and long exposure times (Figure 11). These very high-quality patterns could be used for advanced analysis strategies, such as direct pattern comparison for elastic strain analysis using HR-EBSD and determination of crystal symmetry (Baba-Kishi & Dingley, 1989; Burkhardt et al., 2020). This is important as while patterns captured by this detector only have 256 x 256 pixels, high angular, high frequency features contain rich information about the unit cell. Britton et al. (2013a & 2013b) demonstrate that it is the angular information encoded within the diffraction pattern (i.e. high spatial frequencies in the diffraction pattern and high SNR), rather than the total number of pixels, which is important for the ultimate resolution for high angular resolution EBSD studies. This has been supported more recently by the study of Winkelmann et al. (2025) which suggests that similar pattern sizes may be sufficient for most pattern matching-based EBSD analysis.

Throughout these experiments, we have collected patterns with a pattern center that is towards the top of the EBSD pattern, like the optimal positioning of the pattern center for a conventional EBSD experiment. Note that Zhang & Britton (2024) have previously demonstrated that a DED detector has sufficient SNR, such that EBSD patterns with a wide range of pattern center values can still be indexed.

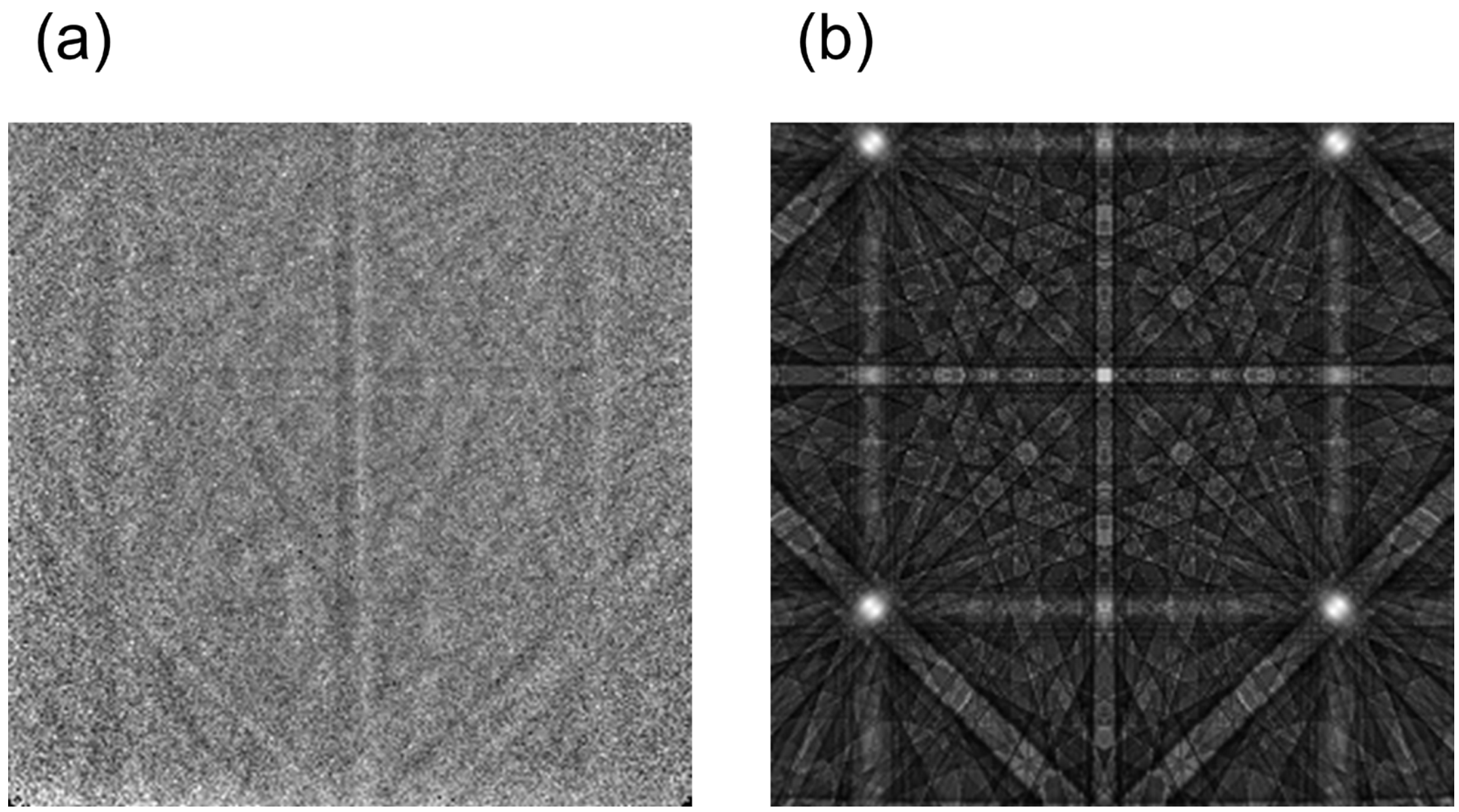

Figure 12. (a) “Detector pattern” at 20 keV and (b) a corresponding dynamical simulation of an EBSD pattern with the same crystal orientation and pattern center.

Below we also address certain practical disadvantages and complications of using similar DEDs for EBSD. A complication associated with the present type of detector is the “detector pattern” that forms when a divergent point source of electrons with approximately uniform illumination strikes the single crystal Si active layer. This forms a detector pattern which is a contrast-inverted Kikuchi pattern, which is formed in a similar fashion as an electron channeling pattern (Vespucci et al., 2017). An example captured with the present setup is shown in Figure 12 alongside a simulation of a dynamical electron channeling pattern of the same crystal orientation and electron source point. Note that the contrast here is inverted, as channeling in the Si-sensor layer decreases the backscatter loss in the sensor layer.

The “detector pattern” is very apparent when probing a non-diffracting region, and it is typically not observed when the diffraction pattern from the sample is of high quality and high signal to noise. For pattern flat fielding using the sum or average of multiple “patterns” from the sample, one can imagine that for a sample with a considerable portion of non-diffracting features, the detector pattern may be a prominent part of the background, and therefore “flat-fielded” patterns may also contain features of the “detector pattern” which could adversely affect indexing. Alternative strategies for background correction include (1) Fourier filtering or a calculated background correction (e.g. a Gaussian mask) for each individual raw diffraction patterns, and (2) collecting the background form a sufficiently large sample area (i.e. > 5 detector pixels, i.e. 250 µm horizontal field width) so that the detector pattern shifts significantly and will be blurred out in the background.

Additionally, the Timepix based detector has a small physical sensor size, which means that the sensor is quite close to the sample for the experiments conducted here (a typically pattern center is [0.484, 0.259, 0.497], using the convention described by Britton et al., (2016)). A detector distance of 0.497 means that the 14 x 14 mm sized detector is only ~7 mm from the sample. To address some challenges that this poses, physical shielding of the detector assembly and built-in safety features inside the microscope control reduce the risk of touching the detector with the sample. In addition, the bias voltage applied to the sensor will also cause changes in the SEM signal as the detector is inserted. To reduce the proximity effect, a viable route is to tile multiple sensors to form a larger detector, and then for the same normalized detector distance the detector will be further away from the sample.

For lower symmetry materials, often a closer sample-to-detector distance/ratio (<0.6) can be suggested so that more zone axes are captured within the diffraction pattern to avoid misindexing due to pseudo symmetry, and this is challenging with a single chip based Timepix sensor. However, there are alternatives possible to address pseudo-symmetry and indexing challenges that are experienced in the analysis of minerals and other ceramics, as the SNR of the patterns collected with a DED are typically higher than that collected by an indirect detector. This means that for the same incident electron dose sharper patterns are likely to be collected and so this is unlikely to be a significant challenge even for users studying most minerals. Furthermore, users have access to pattern matching (Foden et al., 2019; Chen et al., 2015; Winkelmann et al., 2020) or spherical indexing methods (Hielscher et al., 2019; Lenthe et al., 2019) which can also be employed to improve the reliability of pattern indexing.

In addition to these approaches which motivate the optimization of individual diffraction pattern collection to enable high quality and reliable EBSD analysis, users should continually remember that collection of an EBSD map can be viewed as an 'experiment' in its own right. This ethos can encourage a motivated operator to optimize the step size and map scanning strategy to enable the appropriate amount of data in the most efficient manner, saving instrument time and enabling problems to be solved more quickly. We also acknowledge that the application of DEDs to EBSD or Kikuchi diffraction in general has just started. Systematic studies on different direct detection technologies and detectors, for example, the effect of pixel size and pixel array size, can greatly improve the community understanding of these detectors within the context of Kikuchi diffraction.

## Conclusions

This work highlights the effective use of a compact direct electron detector for routine EBSD analysis, as well as outlining the potential for advancing EBSD analysis. A systematic approach studying a single crystal of silicon was used to explore optimal microscope and detector settings, to collect the most useful EBSD patterns for different experiments.

For conventional EBSD analysis, use of a moderate voltage (20 keV) of primary electron beam with a reasonable probe current (12 nA) can result in the collection of microstructure maps very efficiently, with pattern collection and analysis rates that are ~2000 pps. For different sample types, these optimal parameters can be varied, and guidance is provided on how a user may wish to establish these parameters.

For higher pattern qualities, the threshold can be increased to nearer the primary energy and the detected electron dose can be increased by increasing the exposure time and/or primary beam current.

This Timepix based DED can also operate at lower primary beam energies (>5 keV), provided that the sample surface is of very high quality and the beam current, exposure time and beam energy are optimized.

## CRediT Authorship Contribution Statement

T. Zhang: investigation, methodology, data curation, formal analysis, visualization, validation, writing – original draft. R. Birch: investigation, methodology, validation. G. Francolini: investigation, methodology. E. K. Uluscu: investigation, methodology. T. B. Britton: conceptualization, funding acquisition, project administration, resources, supervision, visualization, writing – review & editing.

## Data availability

The data is available via DOI: 10.5281/zenodo. MATLAB code to load and analyze the pattern data, as well for analysis of the microstructure maps within MTEX has been included in the latest release of AstroEBSD (https://github.com/ExpMicroMech/AstroEBSD).

## Acknowledgements

We acknowledge the following funding support: Natural Sciences and Engineering Research Council of Canada (NSERC) [Discovery grant: RGPIN-2022-04762, 'Advances in Data Driven Quantitative Materials Characterization']; British Columbia Knowledge Fund (BCKDF) Canada Foundation for Innovation – Innovation Fund (CFI-IF) [#39798, 'AM+'] and [#43737, 3D-MARVIN]. EKU would like to acknowledge funding from the Republic of Türkiye Ministry of National Education. The authors would like to thank Drs Jakub Holzer, Chris Stephens (Thermo Fisher Scientific) and Kirsty Paton (Paul Scherer Institute) for helpful discussions. The SiC sample was provided by Prof Finn Giuliani (Imperial College London) and the Ti-Mo sample was provided by Dr Mariana Mendes (University of British Columbia).